\documentclass[aps,prl,reprint,twocolumn,showpacs,superscriptaddress]{revtex4-1}
\usepackage{amsmath,amssymb,amsfonts,bm,graphicx,color}

\begin{document}

\title{Heat engine driven by photon tunneling in many-body systems}

\author{Ivan Latella}
\email{ilatella@ffn.ub.edu}
\affiliation{Departament de F\'{i}sica Fonamental, Facultat de F\'{i}sica, Universitat de Barcelona, Mart\'{i} i Franqu\`{e}s 1, 08028 Barcelona, Spain}
\affiliation{Laboratoire Charles Fabry, UMR 8501, Institut d'Optique, CNRS, Universit\'e Paris-Sud 11, 2 Avenue Augustin Fresnel,
91127 Palaiseau Cedex, France}
\author{Agust\'in P\'erez-Madrid}
\affiliation{Departament de F\'{i}sica Fonamental, Facultat de F\'{i}sica, Universitat de Barcelona, Mart\'{i} i Franqu\`{e}s 1, 08028 Barcelona, Spain}
\author{J. Miguel Rubi}
\affiliation{Departament de F\'{i}sica Fonamental, Facultat de F\'{i}sica, Universitat de Barcelona, Mart\'{i} i Franqu\`{e}s 1, 08028 Barcelona, Spain}
\author{Svend-Age Biehs}
\affiliation{Institut f\"ur Physik, Carl von Ossietzky Universit\"at, D-26111 Oldenburg, Germany}
\author{Philippe Ben-Abdallah}
\email{pba@institutoptique.fr}
\affiliation{Laboratoire Charles Fabry, UMR 8501, Institut d'Optique, CNRS, Universit\'e Paris-Sud 11, 2 Avenue Augustin Fresnel,
91127 Palaiseau Cedex, France}


\begin{abstract}
Near-field heat engines are devices that convert the evanescent thermal field supported by a primary source into usable mechanical energy. 
By analyzing the thermodynamic performance of three-body near-field heat engines, we demonstrate that the power they supply can be substantially larger than that of two-body systems, showing their strong potential for energy harvesting.
Theoretical limits for energy and entropy fluxes in three-body systems are discussed and compared with their corresponding two-body counterparts.
Such considerations confirm that the thermodynamic availability in energy-conversion processes driven by three-body photon tunneling can exceed the thermodynamic availability in two-body systems.
\end{abstract}


\maketitle

\newcommand{\dif}{\mathop{}\!\text{d}}
\newcommand{\kB}{k_B}
\newcommand{\vect}[1]{\bm{#1}}

A heat engine may, in general, be conceived as a device that converts part of the heat coming from a hot source of energy into mechanical work throughout an appropriate conversion system~\cite{Callen}.
In contactless devices this heat is transferred to the converter by radiation, only.
At long separation distances the maximum power which can be transmitted is bounded by the 
blackbody limit~\cite{Planck}. On the contrary, at separation distances smaller than the thermal 
wavelength, heat can be transferred to the converter also by photon tunneling, so that the flux 
can become several orders of magnitude larger than in the far-field regime, as shown
theoretically~\cite{Polder} and experimentally~\cite{Kittel,Narayanaswamy1,Narayanaswamy2,Rousseau,Ottens,Kralik}.
Furthermore, it could be shown experimentally and theoretically that near-field thermophotovoltaic conversion 
devices can be used to harvest this energy by transferring it towards a $p$-$n$ junction~\cite{DiMatteo,Narayanaswamy3}. 
Thanks to the tunneling of surface phonon polaritons (SPPs) supported by the primary source, this energy transfer is
quasimonochromatic, which is very advantageous for the energy conversion with a photovoltaic cell~\cite{Narayanaswamy3}.
However, recent theoretical works~\cite{Ivan} have demonstrated the strong potential of near-field heat 
exchanges in the most general context of heat engines for capturing low-grade waste heat for 
power generation. Beside this result, a mechanism of photon-tunneling enhancement has been predicted 
in three-body (3B) systems~\cite{Ben-AbdallahBiehs,Ben-Abdallah2,Ben-Abdallah3,Messina} when passive relays are used to connect 
two bodies in interaction. In this paper, we study the thermodynamic performance of such a 3B 
system and demonstrate the strong potential of these near-field heat engines for energy 
harvesting. 

\begin{figure}
\includegraphics[scale=0.97]{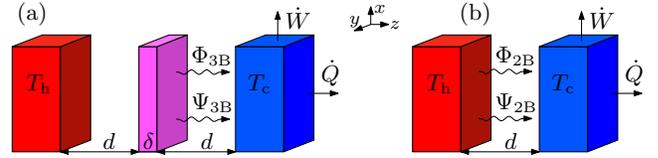}
\caption{Sketch of a heat engine with a hot source at temperature $T_h$ and a cold sink at temperature $T_c<T_h$ that provides a usable work flux $\dot{W}$ by converting near-field thermal radiation energy.
The cold sink receives a heat flux $\dot{Q}$.
(a) Three-body: One of the bodies (emitters) is thermalized with the source and another one with the sink, while a passive intermediate body (of width $\delta$) is placed between them.
The net energy and entropy fluxes on the cold body are $\Phi_\text{3B}$ and $\Psi_\text{3B}$, respectively.
(b) Two-body: The intermediate body is removed.
The net energy and entropy fluxes on the cold body are $\Phi_\text{2B}$ and $\Psi_\text{2B}$, respectively.
The distance $d$ between the bodies is indicated in both cases.}
\label{sketch}
\end{figure}

The properties of the thermal radiation driving an energy-conversion process depend on the distribution and the number of bodies interacting with the converter. Let us compare the operating modes of two-body (2B) and 3B radiative heat engines, which are both
sketched in Fig.~\ref{sketch}. In a 2B heat engine a hot source at temperature $T_h$ radiates towards a 
converter which is assumed to be in contact with a cold sink at temperature $T_c<T_h$. In the 
3B configuration, a passive intermediate body of width $\delta$ is placed between the source and the sink. 
Note that this passive relay is maintained at the same separation distance $d$ from both the source and the 
sink as the cavity width in the 2B system. Hence, we do not introduce in the 3B heat engine an exaltation 
mechanism which results from a simple reduction of distances.
Moreover, we assume that the intermediate body reaches a local equilibrium temperature $T_r$, an assumption that is justified in practical applications for the sizes of the body that we consider here.
This temperature $T_r$ is not arbitrary. It is 
taken such that the net energy flux that the intermediate body exchanges with the hot and cold bodies vanishes. Hence, $T_r$ is an implicit 
function of $T_h$, $T_c$, and the two parameters $d$ and $\delta$ that specify the geometry of the problem.
As a consequence, the energy flux radiated by the hot body coincides with the flux received by the cold body.
This ensures that the energy supplied to the system comes only from the hot source, since, under these conditions, a thermostat at $T_r$ in contact with the passive relay will provide a vanishing net energy flux.

The planar 2B and 3B structures considered here have an infinite transversal extension.
When all the bodies are separated by vacuum, the net energy flux on 
the cold body can be written as an integral over monochromatic contributions of frequency $\omega$, which in 
the near-field regime is given by $\Phi_{i\text{B}}=\int_0^\infty\frac{\dif\omega}{2\pi}\phi_{i\text{B}}(\omega,d,\delta)$ ($i=2,3$) with
\begin{eqnarray}\label{energy3b}
&&\phi_\text{3B}(\omega,d,\delta)=\hbar\omega \sum_j\int_{c\kappa>\omega}\frac{\dif^2\vect{\kappa}}{(2\pi)^2} \\
&&\qquad\times\left[n_{hr}(\omega)\mathcal{T}_j^{(hr)}(\omega,\kappa,d,\delta)+n_{rc}(\omega)\mathcal{T}_j^{(rc)}(\omega,\kappa,d,\delta)\right]\nonumber
\end{eqnarray}
for the 3B configuration~\cite{Ben-Abdallah2} and 
\begin{equation}\label{energy2b}
\phi_\text{2B}(\omega,d)=\hbar\omega \sum_j\int_{c\kappa>\omega}\frac{\dif^2\vect{\kappa}}{(2\pi)^2} n_{hc}(\omega)\mathcal{T}_j^{(hc)}(\omega,\kappa,d)
\end{equation}
in the 2B case~\cite{Polder,Rytov}. Above, we have introduced $n_{\alpha\beta}(\omega)=n_\alpha(\omega)-n_\beta(\omega)$, 
where $n_\alpha(\omega)=\left(e^{\hbar\omega/\kB T_\alpha}-1\right)^{-1}$ are the distributions of photons at 
equilibrium temperature $T_\alpha$ with $\alpha=h,r,c$, $\kB$ being Boltzmann's constant 
and $2\pi\hbar$ Planck's constant. In Eqs.~(\ref{energy3b}) and (\ref{energy2b}), the sum runs over the polarizations $j=s,p$, 
and the integration is carried out over the components of the transverse wave vector $\vect{\kappa}=(k_x,k_y)$  
with $\kappa=|\vect{\kappa}|>\omega/c$, $c$ being the speed of light in vacuum. This means that only the dominant near-field contribution
of evanescent waves is taken into account, whereas the contribution of the propagating modes with $\kappa < \omega/c$
is neglected. The transmission coefficients of the 2B and 3B systems are defined as~\cite{Polder,Rytov,Ben-Abdallah2}
\begin{eqnarray}
&&\mathcal{T}^{(hr)}_j=4\left|\tau^r_j\right|^2 \text{Im}\left(\rho^h_j\right)\text{Im}\left(\rho^c_j\right)e^{- 4|k_z|d}/ \left|D_j^{hrc}D_j^{hr}\right|^2,\nonumber
\\
&&\mathcal{T}^{(rc)}_j=4 \text{Im}\left(\rho^{hr}_j\right) \text{Im}\left(\rho^c_j\right)e^{- 2|k_z|d}/\left|D^{hrc}_j\right|^2,
\label{transmission_coefficients}\\
&&\mathcal{T}^{(hc)}_j=4 \text{Im}\left(\rho^h_j\right) \text{Im}\left(\rho^c_j\right)e^{- 2|k_z|d}/\left|D^{hc}_j\right|^2,\nonumber
\end{eqnarray}
where $k_z=\sqrt{\omega^2/c^2-\kappa^2}$ is the normal component of the wave vector while 
$D^{hr}_j=1- \rho^h_j\rho^r_je^{2ik_zd}$,  $D^{hrc}_j=1- \rho^{hr}_j\rho^c_je^{2ik_zd}$ 
and  $D^{hc}_j=1- \rho^h_j\rho^c_je^{2ik_zd}$  are the Fabry-P\'erot--like denominators.
Here $\rho^h_j=\rho^h_j(\omega,\kappa)$, $\rho^r_j=\rho^r_j(\omega,\kappa,\delta)$, 
and $\rho^c_j=\rho^c_j(\omega,\kappa)$ are the reflection coefficients of the hot, intermediate, 
and cold bodies, respectively, $\tau^r_j=\tau^r_j(\omega,\kappa,\delta)$ are the transmission 
coefficients of the intermediate body, and $\rho^{hr}_j=\rho^r_j+\left(\tau^r_j\right)^2\rho^h_je^{2i k_zd}/D_j^{hr}$ 
are the reflection coefficients of the hot and the intermediate bodies considered as a single entity.

\begin{figure}
\includegraphics[scale=1]{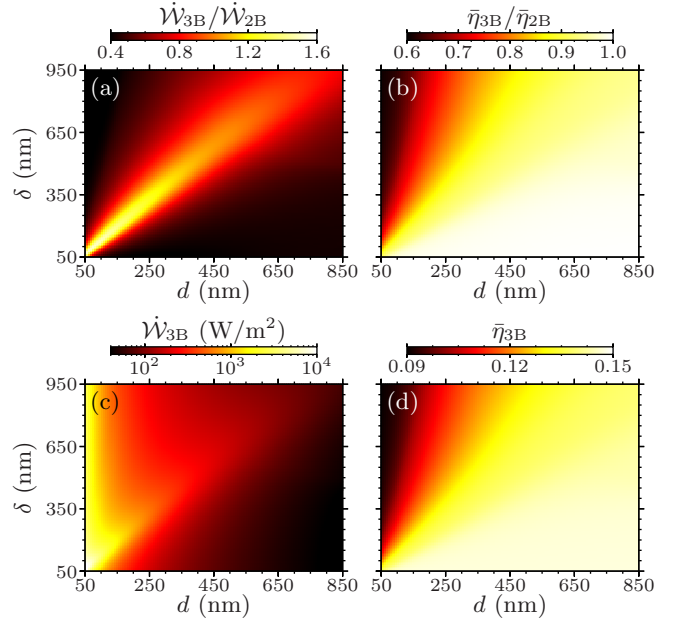}
\caption{(a) Ratio of the maximum work flux in the 3B configuration to the maximum work flux in the 2B configuration, $\dot{\mathcal{W}}_\text{3B}/\dot{\mathcal{W}}_\text{2B}$, as a function of the separation $d$ and width of the intermediate body $\delta$ for $T_h=400\,$K and $T_c=300\,$K.
A region of amplification due to 3B photon tunneling is clearly appreciated. (b) Efficiency ratio $\bar{\eta}_\text{3B}/\bar{\eta}_\text{2B}$ in the same conditions.
In (c) and (d), the plots show the corresponding maximum work flux and efficiency in the 3B configuration.}
\label{ratios}
\end{figure}

The expressions (\ref{transmission_coefficients}) for the transmission coefficients show that the three bodies are 
coupled together due to multiple interaction mechanisms resulting in their nontrivial optical properties.
Moreover, in view of Eqs.~(\ref{energy3b}) and (\ref{energy2b}),  the energy flux $\Phi_{i\text{B}}$ can be written as a sum 
over the contributions stemming from the different constituents of the system which are in local thermal equilibrium, 
i.e.\ $\Phi_{i\text{B}}=\sum_\alpha\Phi^{(i\text{B})}_\alpha(T_\alpha)$ with $\alpha=h,c$ for 2B 
and $\alpha=h,r,c$ in the 3B case. The fluxes $\Phi^{(i\text{B})}_\alpha(T_\alpha)$ depend only 
on the local equilibrium temperatures $T_\alpha$ of the constituents through the distribution functions $n_\alpha(\omega)$.
Hence, the partial entropy fluxes $\Psi^{(i\text{B})}_\alpha(T_\alpha)$ carried by the thermal fields generated by the different
constituents are given by~\cite{Ivan}
\begin{equation}
\Psi^{(i\text{B})}_\alpha(T_\alpha)=\int_0^{T_\alpha} \!\! \dif T' \,\frac{1}{T'}\frac{\dif}{\dif T'}\Phi^{(i\text{B})}_\alpha(T'). 
\label{entropy_fluxes}
\end{equation}
Therefore, the net entropy flux on the cold body reads
\begin{equation}
\Psi_{i\text{B}}=\sum_\alpha\Psi^{(i\text{B})}_\alpha(T_\alpha)= \int_0^\infty\frac{\dif\omega}{2\pi}\psi_{i\text{B}}(\omega,d,\delta),
\end{equation}
where the spectral entropy fluxes take the form
\begin{eqnarray}
&&\psi_\text{3B}(\omega,d,\delta)=\kB \sum_j\int_{c\kappa>\omega}\frac{\dif^2\vect{\kappa}}{(2\pi)^2} \label{entropy3b}\\
&&\quad\ \times\left[m_{hr}(\omega)\mathcal{T}_j^{(hr)}(\omega,\kappa,d,\delta)
+m_{rc}(\omega)\mathcal{T}_j^{(rc)}(\omega,\kappa,d,\delta)\right]\nonumber,\\
&&\psi_\text{2B}(\omega,d)=\kB\sum_j\int_{c\kappa>\omega}\frac{\dif^2\vect{\kappa}}{(2\pi)^2}m_{hc}(\omega)\mathcal{T}_j^{(hc)}(\omega,\kappa,d), 
\label{entropy2b}
\end{eqnarray}
with $m_{\alpha\beta}(\omega)=m_\alpha(\omega)-m_\beta(\omega)$ and $m_\alpha(\omega)=\left[1+n_\alpha(\omega)\right]\ln\left[1+n_\alpha(\omega)\right]-n_\alpha(\omega)\ln n_\alpha(\omega)$. These last two relations are strictly valid only if the temperature dependence of the material
properties of the constituents of the 2B or 3B system can be neglected in the considered range of working temperatures.

Once energy and entropy fluxes are known, the thermodynamics of the energy-conversion process can be analyzed as 
follows (the 3B and the 2B configurations will be discussed simultaneously). First of all, notice that, due to the 
difference of temperatures between the bodies, the transport of heat through the cavity proceeds irreversibly and entropy 
is generated at a certain rate, say, $\Psi_g$.
This entropy production $\Psi_g$ accounts for dissipative processes in the thermalization of excited electrons at the surface of the cold body~\cite{DeVos}.
Since the bodies are thermalized, in particular, a heat 
flux $\dot{Q}$ is transferred isothermally to the cold sink; we assume that this transference is done reversibly and, 
thus, $\dot{Q}=T_c\left(\Psi_{i\text{B}}+\Psi_g\right)$.
In this scheme the heat engine can be considered as endoreversible, as discussed in~\cite{DeVos} for the 2B problem in the far field.
Taking into account the balances of energy and entropy fluxes, the 
work flux that can be delivered by the engine reads $\dot{W}=\Phi_{i\text{B}}-T_c\left(\Psi_{i\text{B}}+\Psi_g\right)$.
Since $\Psi_g\geq0$, the maximum work flux or thermodynamic availability is given 
by $\dot{\mathcal{W}}_{i\text{B}}\equiv \Phi_{i\text{B}}-T_c\Psi_{i\text{B}}$. In addition, considering 
$\Phi_{i\text{B}}$ as the input energy flux, the efficiency of the engine is given by $\eta_{i\text{B}}=\dot{W}/\Phi_{i\text{B}}$.
According to this, an upper bound for the efficiency can be obtained by computing the 
ratio $\bar{\eta}_{i\text{B}}\equiv\dot{\mathcal{W}}_{i\text{B}}/\Phi_{i\text{B}}$.

In the configuration we analyze, the hot and cold bodies are two 5-$\mu$m-thick silicon carbide (SiC) samples~\cite{Palik} that support a SPP with a resonance at $\omega_\text{SPP}\simeq 1.79\times10^{14}\,$rad/s.
Moreover, as in Ref.~\cite{Ben-Abdallah2}, we use in the 3B configuration for the intermediate slab a metallike medium which supports a surface 
mode (a plasmon) at the same frequency $\omega_\text{SPP}$.
Figure~\ref{ratios}(a) shows the ratio of the maximum work flux in the 3B system to the maximum work flux in the 2B system, i.e., with and without an intermediate relay.
It can be seen that  a 3B engine can produce about $60\%$ more work 
than a classical 2B system. If the width of the intermediate body becomes sufficiently large, the 3B interaction 
disappears (in the near-field regime), and both cavities, located between the source and the intermediate relay and 
between the relay and the sink, become independent. Then, the work production by the 3B heat engine becomes comparable to or even smaller than the one of a 2B engine. As for the efficiency of those engines, we see in Fig.~\ref{ratios}(b) that they 
are comparable in both configurations provided the separation distances are large enough compared to the width of the 
intermediate slab. It is interesting to note that the 2B efficiency seems always to be larger than the 3B
efficiency even in the parameter range where the extracted work of the 3B system exceeds that of the 2B system.

\begin{figure}
\includegraphics[scale=1]{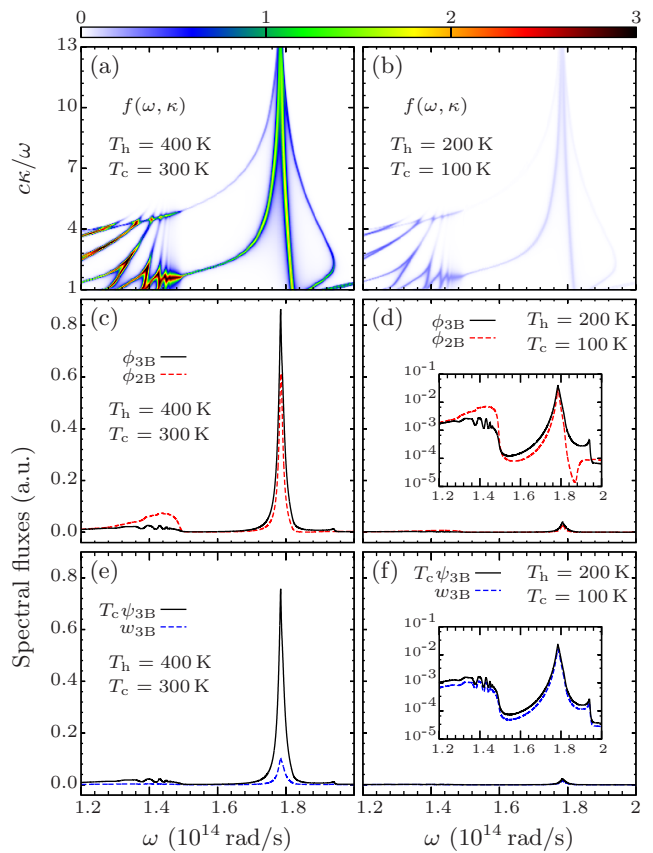}
\caption{(a) Transmission coefficients weighted by the photon distributions taking $d=500\,$nm and $\delta=667\,$nm.
We plot $f(\omega,\kappa)=10^{22}\times\left(n_{hr}\mathcal{T}_p^{(hr)}+n_{rc}\mathcal{T}_p^{(rc)}\right)$ for $T_h=400\,$K and $T_c=300\,$K, for which $T_r=357.01\,$K.
In (b), $f(\omega,\kappa)$ is shown for $T_h=200\,$K and $T_c=100\,$K with $T_r=180.54\,$K.
(c) Spectral energy fluxes $\phi_\text{3B}$ and $\phi_\text{2B}$ in 3B and 2B configurations, respectively, corresponding to the same setting used in (a).
(d) Spectral energy fluxes $\phi_\text{3B}$ and $\phi_\text{2B}$ corresponding to (b); the inset shows the same spectra in log scale.
In (e) and (f) we plot the spectral entropy flux (multiplied by the temperature of the sink) $T_c\psi_\text{3B}$ and the spectral work flux $w_\text{3B}=\phi_\text{3B}-T_c\psi_\text{3B}$ corresponding to the same setting used in (a)--(c) and in (b)--(d), respectively.
}
\label{spectra}
\end{figure}

In order to get some insight on these results, we plot in Figs.~\ref{spectra}(a) and \ref{spectra}(b) the transmission coefficients 
for $p$-polarized waves (the main contribution) in the ($\omega,\kappa$) plane associated to the 3B engine by 
weighting them with the corresponding photon distribution functions. As a first observation, we see the presence 
of different surface-mode branches of the four coupled surface modes (symmetric and antisymmetric modes) in 
the 3B system~\cite{Economou1969} around the surface-mode resonance frequency of SiC. For these surface modes,
the transmission is apparently high. These branches support high transmissions for large wave vectors, which means 
that a large number of modes contribute to the heat transfer in this spectral region~\cite{Ben-Abdallah1,BiehsEtAl2010}. 
The closer the frequency of the surface mode gets to Wien's frequency of the heat source $\omega_W=2.82\kB T_h/\hbar$, 
the higher the number of excitations that contribute to the transfer. Accordingly, if the hot body is cooled down 
to a temperature for which $\omega_W$ is far from $\omega_\text{SPP}$, the modes in the region around the 
SPP stop to contribute to the transfer, as can be seen in Fig.~\ref{spectra}(b). The spectral energy fluxes plotted 
in Fig.~\ref{spectra}(c) and \ref{spectra}(d) corroborate this tendency. Since the 3B photon-tunneling enhancement occurs in the SPP region, we thus observe in Fig.~\ref{spectra}(c) an increase 
of the quasimonochromatic spectral energy flux $\phi_\text{3B}$ as compared with $\phi_\text{2B}$.
Furthermore, the spectral entropy flux $\psi_\text{3B}$ and the spectral work flux $w_\text{3B}\equiv\phi_\text{3B}-T_c\psi_\text{3B}$ are also peaked around the SPP frequency. As shown in Fig.~\ref{spectra}(e), the negative entropic term drastically reduces the monochromatic contribution $w_\text{3B}$ to the thermodynamic availability $\dot{\mathcal{W}}_{3\text{B}}(d,\delta)=\int_0^\infty\frac{\dif\omega}{2\pi}w_{3\text{B}}(\omega,d,\delta)$. This entropic term represents a non-negligible energy flux that the system transfers to the cold sink, thus diminishing the amount of usable work production.

\begin{figure}
\includegraphics[scale=1]{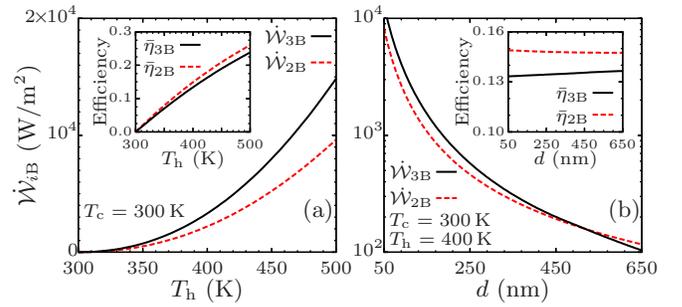}
\caption{(a) Maximum work fluxes $\dot{\mathcal{W}}_\text{3B}$ and $\dot{\mathcal{W}}_\text{2B}$ for the 3B and 2B configurations, 
respectively, as a function of the temperature of the hot source $T_h$. The temperature of the cold sink is 
set to $T_c=300\,$K, the separation distance to $d=100\,$nm, and the width of the intermediate body to $\delta=133\,$nm.
The inset shows the corresponding upper bounds for the efficiency, $\bar{\eta}_\text{3B}$ and $\bar{\eta}_\text{2B}$.
(b) Dependence of the maximum work fluxes and efficiencies on the separation distance $d$ for fixed temperatures and the optimal $\delta$.}
\label{fluxes}
\end{figure}

Finally, we study the maximal work flux $\dot{\mathcal{W}}_{i\text{B}}$ that can be extracted from such 
a 3B heat engine compared with that of a 2B heat engine. To this end, we fix $T_c=300\,$K, while the temperature of the heat
source $T_h$ is varied. For the thickness of the vacuum gaps and the intermediate passive relay
we choose $d = 100\,{\rm nm}$ and $\delta=133\,$nm for which $\dot{\mathcal{W}}_\text{3B}/\dot{\mathcal{W}}_\text{2B}$
is maximum when $T_h$ has reached a temperature of $400\,{\rm K}$, as shown in Fig.~\ref{ratios}(a) (in that case, the ratio $\dot{\mathcal{W}}_\text{3B}/\dot{\mathcal{W}}_\text{2B}$ slowly increases for increasing $T_h$).
The results are plotted in Fig.~\ref{fluxes}(a). They show that the discrepancy 
between $\dot{\mathcal{W}}_\text{3B}$ and $\dot{\mathcal{W}}_\text{2B}$ grows monotonically with respect 
to temperature while the 2B and 3B efficiencies remain very close to each other 
($\overline{\eta}_{\rm 2B} \geq \overline{\eta}_{\rm 3B}$).
In addition, the dependence of the work fluxes and efficiencies on the distance $d$ for the optimal $\delta$ are presented in Fig.~\ref{fluxes}(b),
where the temperatures are set to $T_h=400\,$K and $T_c=300\,$K.
This example illustrates that in a 3B system
the energy flux and the  maximal work flux that can be extracted are enhanced by the interactions
of the surface modes in the hot and cold body with that of the intermediate relay. 

The maximum transfer in a 3B configuration takes place when the transmission coefficients attain their maximum value. The theoretical limit is thus achieved by the condition $\mathcal{T}_j^{(hr)}=\mathcal{T}_j^{(rc)}=1$, as also occurs for 2B systems~\cite{Pendry,Volokitin2004,Ben-Abdallah1} when $\mathcal{T}_j^{(hc)}=1$.
Using this in (\ref{energy3b}), (\ref{energy2b}), (\ref{entropy3b}), and (\ref{entropy2b}) and taking into account a cutoff wave vector $\kappa_{c,i\text{B}}\gg\omega/c$, for which the modes are effectively confined, we get $\Phi_{i\text{B}}^{\rm max} = \xi_{i\text{B}} \left(T_h^2-T_c^2\right)$ and $\Psi_{i\text{B}}^{\rm max} = 2\xi_{i\text{B}} \left(T_h-T_c\right)$, where $\xi_{i\text{B}} = \kappa_{c,i\text{B}}^2\kB^2/24\hbar$.
Notice that maximizing the energy flux implies also that the flow of entropy per channel is maximum~\cite{Pendry}.
The maximum work flux is thus $\dot{\mathcal{W}}_{i\text{B}}^{\rm max}=\xi_{i\text{B}}(T_h-T_c)^2$,
and in consequence the upper bound for the efficiency reads $\bar{\eta}_\text{2B}^{\rm max} = \bar{\eta}_\text{3B}^{\rm max} = (T_h-T_c)/(T_h+T_c)$.
Therefore, the efficiencies for the 2B and 3B systems are equal. However, we remark that the difference between a 3B and 
a 2B system is manifested through $\kappa_{c,i\text{B}}$. The cutoff wave vector in a 3B system can be larger 
than that of the 2B configuration as shown in our numerical examples. Although the efficiencies and the ratios $\Phi_{i\text{B}}^\text{max}/\Psi_{i\text{B}}^\text{max}$ are the same in the 3B and 2B system, it follows that
\begin{equation}
  \frac{\dot{\mathcal{W}}_\text{3B}^{\rm max}}{\dot{\mathcal{W}}^{\rm max}_\text{2B}} = \frac{\Phi_\text{3B}^{\rm max}}{\Phi_\text{2B}^{\rm max}} = \frac{\kappa_{c,\text{3B}}^2}{\kappa_{c,\text{2B}}^2} \geq 1.
\end{equation}
Hence, a larger maximum work flux in the 3B system is due to the larger energy flux which, in turn, results from the larger 
number of contributing modes. 

For the general case of an $N$-body ($N$B) engine, we can always write the heat flux exchanged between the different parts of the system by using the Landauer formalism derived in Refs.~\cite{Ben-Abdallah1,BiehsEtAl2010,Pendry}, the transmission coefficients being simply related to the scattering of the electromagnetic field radiated by each part of the system.
In this case, a suitable cutoff $\kappa_{c,N\text{B}}$ has to be considered. When $\kappa_{c,N\text{B}}>\kappa_{c,2\text{B}}$ for $N>2$, an enhancement in the maximum work flux $\dot{\mathcal{W}}_{N\text{B}}^{\rm max}$ is thus expected with respect to the 2B case, while the efficiency $\bar{\eta}_{N\text{B}}^{\rm max}= (T_h-T_c)/(T_h+T_c)$ remains the same for all $N$.
Furthermore, the heat conductance in this $N$B endoreversible engine is, in general, nonlinear in the difference of temperatures $\Delta T=T_h-T_c$.
However, the linear regime can be achieved if $\Delta T\ll T_h$, and, under this assumption, the efficiency becomes $\bar{\eta}_{N\text{B}}^{\rm max} = \Delta T/(2T_h)$, which is half the Carnot efficiency.
For engines with linear heat conductance such as the Novicov engine, the efficiency is given by $\eta_\text{CA}=1-\sqrt{T_c/T_h}$~\cite{Hoffmann1997}, which for $\Delta T\ll T_h$ becomes $\eta_\text{CA}=\Delta T/(2T_h)$.
We thus observe that the efficiency of our engine, in the appropriate limit, coincides with that of a Novicov engine.

In conclusion, we have demonstrated that the thermodynamic performance of 3B near-field heat engines can substantially 
overcome that of 2B systems. Our results pave the way for a generation of nanoscale energy converters driven 
by the physics of many-body interactions instead of the conventional two-body interactions.
In addition, this work provides perspectives for investigating the thermodynamics of systems with long-range electromagnetic interactions. 

This work was partially supported by the Spanish Government under Grant No. FIS2011-22603.
I.L. acknowledges financial support through an FPI Scholarship (Grant No. BES-2012-054782) from the Spanish Government.
J.M.R. acknowledges financial support from Generalitat de Catalunya under program ICREA Academia.
P.B.-A. acknowledges financial support from the CNRS Energy program.

\end{document}